\documentclass[12pt,preprint]{aastex}

\shorttitle{Detecting Oceans on Extrasolar Planets}
\shortauthors{Robinson et al.}

\begin{document}

\title{Detecting Oceans on Extrasolar Planets Using the Glint Effect}

\author{Tyler D. Robinson\altaffilmark{1}; Victoria S. Meadows\altaffilmark{1}}
\affil{Astronomy Department, University of Washington, Seattle, WA 98195}
\email{robinson@astro.washington.edu}

\and

\author{David Crisp\altaffilmark{1}}
\affil{Jet Propulsion Laboratory, Pasadena, CA 91109, USA}

\altaffiltext{1}{NASA Astrobiology Institute}

\begin{abstract} 
Glint, the specular reflection of sunlight off Earth's oceans, 
may reveal the presence of oceans on an extrasolar planet. As an 
Earth-like planet nears crescent phases, the size of the ocean 
glint spot increases relative to the fraction of illuminated disk, 
while the reflectivity of this spot increases. Both effects change 
the planet's visible reflectivity as a function of phase. However, 
strong forward scattering of radiation by clouds can also produce 
increases in a planet's reflectivity as it approaches crescent 
phases, and surface glint can be obscured by Rayleigh scattering and 
atmospheric absorption. Here we explore the detectability of glint 
in the presence of an atmosphere and realistic phase-dependent 
scattering from oceans and clouds. We use the NASA Astrobiology 
Institute's Virtual Planetary Laboratory 3-D line-by-line, 
multiple-scattering spectral Earth model to simulate Earth's 
broadband visible brightness and reflectivity over an orbit. Our 
validated simulations successfully reproduce phase-dependent 
Earthshine observations. We find that the glinting Earth can be as 
much as 100\% brighter at crescent phases than simulations that do 
not include glint, and that the effect is dependent on both orbital 
inclination and wavelength, where the latter dependence is caused by 
Rayleigh scattering limiting sensitivity to the surface. We show 
that this phenomenon may be observable using the James Webb 
Space Telescope (JWST) paired with an external occulter.
\end{abstract}


\section{Introduction}

A major goal in the study of extrasolar planets is 
the detection and recognition of a ``habitable" world, 
or a planet capable of maintaining liquid water on 
its surface.  A variety of direct and indirect approaches 
could be used to determine if a planet is habitable.  
Indirect approaches focus on characterizing the surface 
environment of a planet, which would constrain the 
likelihood that the planet could maintain liquid water 
on its surface.  Direct approaches aim to detect signs 
that indicate the presence of water on the surface 
of a planet \citep[e.g.,][]{cowanetal09}.
 
One direct indicator of surface bodies of water is specular 
reflection, or 
the ``glint effect".  While specular reflection 
is not unique to liquid surfaces  
\citep[e.g.,][]{dumontetal10}, liquids 
are distinguished from other surfaces 
by their contrast between weak 
specular reflectance at direct illumination 
angles and strong specular reflectance at 
glancing illumination angles.  Recently, 
\citet{stephanetal10} used the enhanced reflectivity 
of liquids at glancing illumination angles to 
provide evidence for liquid hydrocarbon 
lakes on the surface of Titan.

\citet{Saganetal93} argued that the presence of 
a specularly-reflecting region (or ``glint 
spot") in spatially-resolved images of 
Earth taken by the Galileo spacecraft, combined 
with detections of atmospheric water vapor and 
surface temperatures near the melting point of 
water, was evidence for the presence of 
liquid water oceans on Earth's surface.  
Unfortunately, obtaining spatially-resolved images 
of terrestrial extrasolar planets presents an 
engineering challenge that will not be met in the 
near future.  The first measurements that aim 
to detect glint must rely on how it affects the 
brightness of a planet in a disk-integrated 
sense.

The relative size of Earth's glint spot compared to 
the illuminated portion of the disk increases at 
crescent phases, and the reflectivity of water increases 
at glancing illumination angles, affecting Earth's phase 
curve.  The detectability of this effect  
was first investigated by \citet{williams&gaidos08}, 
who used a simple model of Earth's reflectance 
coupled to a 3-D climate model to predict  
Earth's appearance over the course of an orbit. 
Their model showed 
that Earth's reflectivity increases into 
crescent phases, but was unable to reproduce 
Earthshine observations of Earth's reflectivity 
\citep{palleetal03}.  This discrepancy was 
attributed to the absence of Rayleigh scattering 
in their model and the assumption that clouds reflect 
isotropically (Lambertian).  In reality, 
liquid droplets and ice crystals preferentially 
scatter light in the forward direction, which 
can mimic the glint effect.

\citet{oakley&cash09} modeled  
Earth's brightness over the course of an  
orbit with an emphasis on characterization by the New 
Worlds Observer mission concept \citep{cash06}.  
Their model simulated Earth's reflectivity using 
satellite-measured bidirectional 
reflectance distribution functions (BRDFs) 
for a variety of scenes (\emph{e.g.}, thick cloud 
over ocean) \citep{manalosmithetal98}, and used satellite 
observations to evolve clouds, snow, and ice in 
their simulations.  Their model also demonstrated an increase 
in Earth's reflectivity into crescent phases, but was 
not compared to the phase-dependent Earthshine 
measurements. The authors proposed that the bright glint 
spot increases the variability in Earth's brightness at 
crescent phases, which could serve as an indicator of 
surface oceans. However, the BRDFs used in their model 
were not valid at extreme crescent phases, requiring 
assumptions about cloud scattering and ocean reflectivity 
at glancing illumination angles.

In this work we use the NASA Astrobiology Institute's 
Virtual Planetary Laboratory (NAI-VPL) 3-D spectral Earth 
model to simulate Earth's disk-integrated spectrum 
as it would appear to a distant observer watching the 
planet through an orbit.  While extensively 
validated against time-dependent data in 
\citet{robinsonetal10}, we present further 
validations here against phase-dependent 
Earthshine observations.  The validated model is 
used to investigate the significance of glint 
in Earth's phase curve by discriminating between 
the competing effects of cloud scattering and glint. We 
also discuss the observing requirements for glint 
detection.


\section{Model Description}

The NAI-VPL 3-D spectral Earth model simulates 
Earth's appearance to a distant observer.  The 
model produces spatially- and spectrally-resolved datacubes 
that can be collapsed to produce high-resolution, 
disk-integrated spectra, and was 
described and validated in \citet{robinsonetal10}, so 
only a brief summary will be presented here.  Additional 
information can be found in 
\citet{tinettietal06a,tinettietal06b}.

The 3-D spectral Earth model simulates 
Earth's spectrum at arbitrary viewing 
geometry over wavelengths from the far-ultraviolet to the 
far-infrared on timescales from minutes to years. 
Earth-observing satellites provide spatially-resolved, 
date-specific inputs of key surface and atmospheric 
properties.  To simulate 
the seasonal changes in Earth's appearance over a year, we 
use snow cover and sea ice data as well as cloud cover and
optical thickness data from the Moderate Resolution Imaging 
Spectroradiometer (MODIS) instruments 
\citep{salomonsonetal89} aboard NASA's Terra and Aqua 
satellites \citep{halletal95,riggsetal95}.

\begin{figure}[!b]
  \begin{center}
    \includegraphics[width=6in,angle=0]{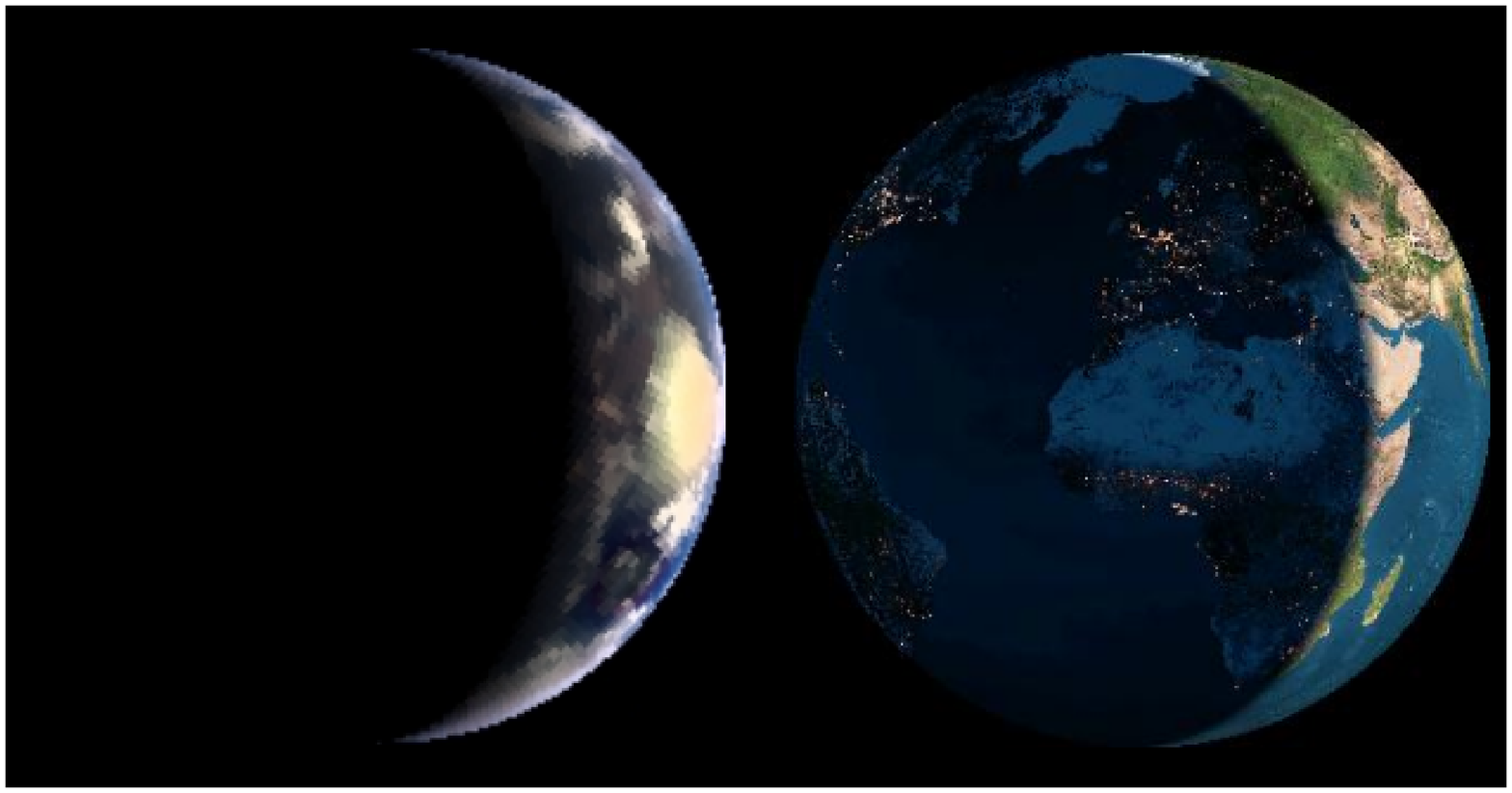}
  \end{center}

  \caption{\footnotesize
           A true-color image from our model (left) 
           compared to a view of Earth from the Earth and 
           Moon Viewer (http://www.fourmilab.ch/cgi-bin/Earth/). A glint spot 
           in the Indian Ocean can be clearly seen in the model image.}
  \label{fig_truecolor}
\end{figure}

Specular reflectance from liquid water surfaces in 
our model is simulated using the Cox-Munk glint 
model \citep{coxmunk54}, which allows for the 
calculation of the BRDF 
of a wave-covered ocean given wind speed 
and direction, which are provided by the QuikSCAT satellite 
(http://winds.jpl.nasa.gov/missions/quikscat/index.cfm).  
Wavelength-dependent optical properties for liquid clouds 
are derived using a Mie theory model \citep{crisp97} and ice 
clouds are parametrized using geometric optics 
\citep{muinonenetal89}.  Figure \ref{fig_truecolor} 
shows a true-color image generated by our model, 
demonstrating the glint effect and the 
ability of clouds to mimic this effect.

\section{Model Validation}

Our model was previously validated against time-dependent 
ultraviolet through mid-infrared observations 
\citep{robinsonetal10}, but for a narrow range of 
phases. Here we extended our validation by matching 
Earthshine measurements of Earth's 
apparent albedo at a variety of phases between gibbous 
and crescent \citep{palleetal03}.  The apparent albedo 
is defined as the albedo of a perfect Lambert sphere 
that would give the same reflectivity of a body at a 
given phase angle. Thus, the apparent albedo of a 
Lambert sphere would be constant through all phases.

Figure \ref{fig_validation} shows our simulation 
of Earth's brightness and corresponding apparent 
albedo as it would appear over the course of 
one year to a distant observer (black curves).   
Observations from NASA's EPOXI mission 
\citep{livengoodetal08} as well as a large number 
of ground-based Earthshine observations are also shown.  
Observations and model results span the wavelength 
range 0.4-0.7 $\mu$m.  EPOXI 
observations were recorded over three separate 
24-hour periods in March, May, and June of 2008, 
while Earthshine measurements span November 
1998 through January 2005. Calibration errors 
as large as 10\% are present in the EPOXI 
observations \citep{klaasenetal08} while Earthshine 
observations have a stated accuracy of 2\% 
\citep{qiuetal03}.

In general, there is good agreement between the 
data and the model, 
demonstrating the ability of the model to properly 
simulate Earth's phase-dependent brightness. At 
gibbous phases where Earthshine data were recorded, the 
mean apparent albedo of the model Earth is $0.29 \pm 0.02$ and 
is $0.26 \pm 0.03$ for the Earthshine data.  Thus, while 
the Earthshine data are systematically lower than the 
EPOXI observations and the model at gibbous phases, they 
are all in agreement to within one standard deviation.  At 
crescent phases the Earthshine observations are 
systematically larger than the model.  However, an 
analysis where the Earthshine data were divided into 
10$^{\circ}$-wide bins in orbital longitude shows that the 
model is always within a single standard deviation of the 
observations.

\section{Results}

\begin{figure}[!b]
  \begin{center}
    \includegraphics[width=5in,angle=0]{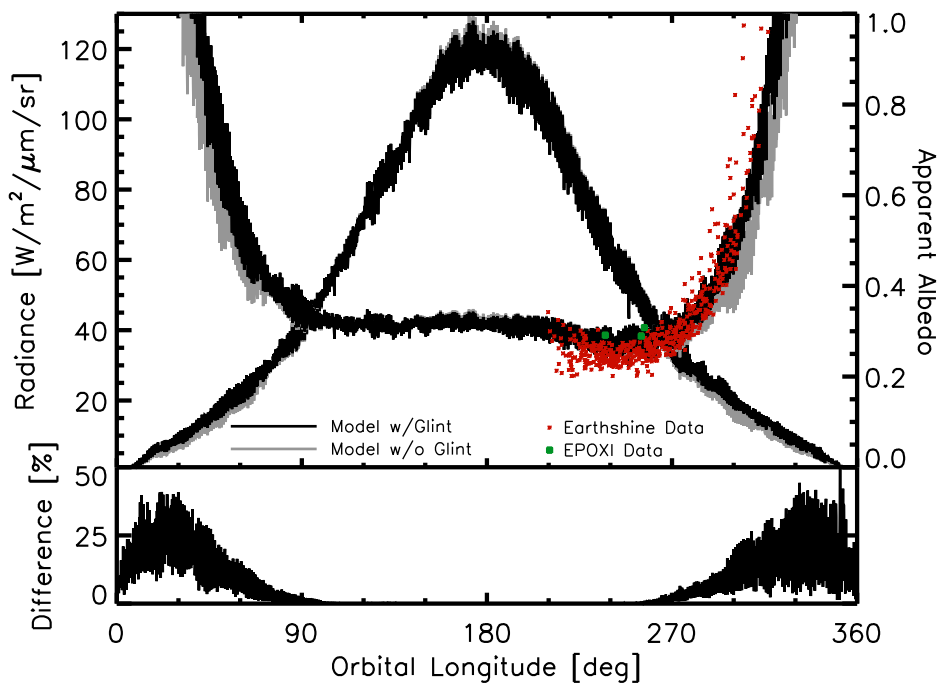}
  \end{center}

  \caption{\footnotesize
           Simulation of Earth through a year.  Black 
           line corresponds to a model that includes glint while the 
           grey line corresponds to a model that does not include 
           glint. The left y-axis corresponds to the bell-shaped 
           curves, demonstrating that Earth is brightest at full 
           phase (orbital longitudes near 180$^{\circ}$) and faintest 
           near crescent phase (orbital longitudes near 0 and 
           360$^{\circ}$).  The right y-axis corresponds to the 
           bowl-shaped curves, where a perfect Lambert sphere 
           would have a constant apparent albedo with phase. 
           Variability at small time scales is due to 
           Earth's rotation and time-varying cloud 
           formations (noise is not included in simulations).  
           Model ``observations" are recorded every four 
           hours, the system is viewed edge-on ($i=90^{\circ}$), and an 
           orbital longitude of 0$^{\circ}$ corresponds to January 1, 
           2008.  Small stars are Earthshine measurements of 
           Earth's apparent albedo \citep{palleetal03}.  
           Large circles are 24-hour average measurements of Earth's 
           apparent albedo recorded by the Deep Impact flyby 
           spacecraft as part of NASA's EPOXI mission \citep{livengoodetal08}. 
           All data and model observations the wavelength range 
           0.4-0.7 $\mu$m. The bottom sub-panel 
           demonstrates the brightness excess seen in the glinting model   
           over the non-glinting model.}
  \label{fig_validation}
\end{figure}

In the subpanel of Fig. \ref{fig_validation}, the glinting  
model demonstrates an excess in 
brightness as large as 50\% over the non-glinting model  
(grey curves), which uses an isotropically-scattering ocean 
reflectance model whose albedo reproduces Earth's geometric 
albedo to within a few percent.  The excess brightness 
increases into crescent phases as the contribution 
of the glint spot to Earth's disk-integrated brightness 
grows, peaking at orbital longitudes near 20-30$^{\circ}$ 
from new phase. The brightness excess declines as the 
illuminated crescent shrinks further.  This 
is due to a wave-surface ``hiding'' effect 
discussed in \citet{coxmunk54}, where ocean waves block 
rays of light at glancing illumination angles.

The brightness difference between the glinting and 
non-glinting model is a strong function of wavelength. 
If we instead choose a filter that spans 
1.0-1.1 $\mu$m, the peak brightness excess increases 
to about 100\%.  This behaviour arises because glint 
occurs at the surface and some wavelength 
ranges are more sensitive to Earth's surface than 
others.

Near 90$^{\circ}$ orbital longitude, both models in 
Fig. \ref{fig_validation} have apparent albedos that 
are about 15-20\% larger than those near 
270$^{\circ}$.  This is due to Earth's 
seasons, as was noted in \citet{williams&gaidos08}, 
and was determined by comparing to a model run without any 
seasonal evolution of snow and ice.  
As Earth moves from northern winter to northern 
summer (0$^{\circ}$ to 180$^{\circ}$ orbital 
longitude in these simulations, respectively), 
the illumination of the northern polar region, which 
is tilted towards the observer, increases.  Since this 
region of Earth is more 
snow- and ice-covered prior to northern summer than 
following northern summer, the planet appears more 
reflective heading into full phase 
than moving out of full phase.  The magnitude of 
this asymmetry in the lightcurve agrees with the 
simulations of \citet{oakley&cash09}.  In general, 
the effects of seasons on Earth's lightcurve 
depend on viewing geometry, but are small 
compared to the effects of cloud scattering 
and glint at crescent phases.

\begin{figure}[!b]
  \begin{center}
    \includegraphics[width=6in,angle=0]{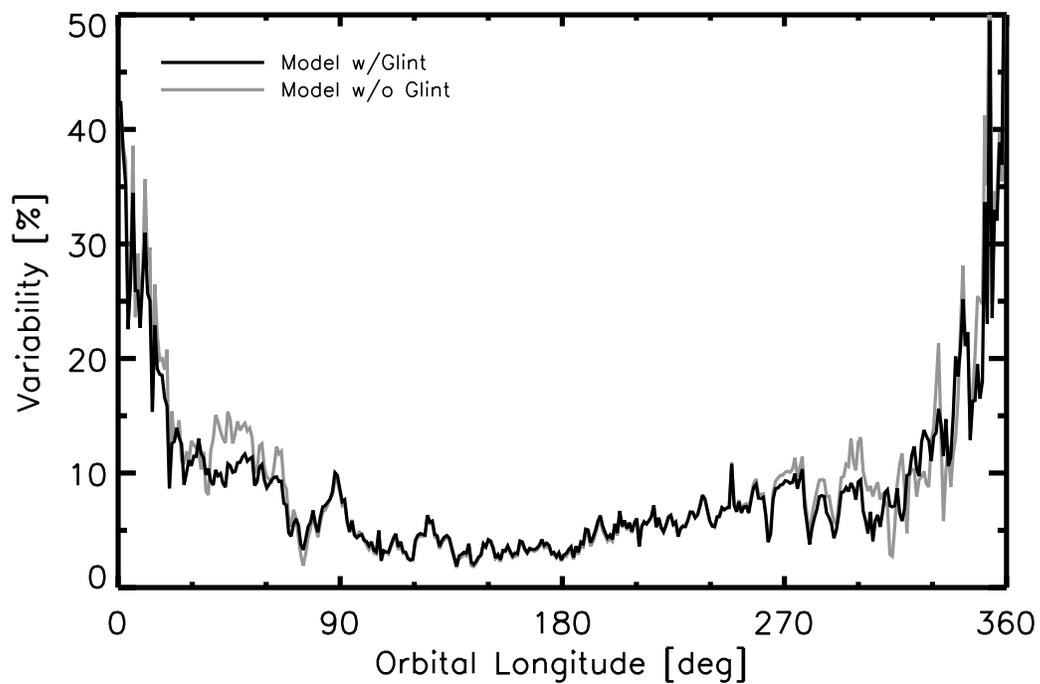}
  \end{center}

  \caption{\footnotesize
           Variability in brightness for our glinting model  
           (black) and our non-glinting model (grey), which 
           are both shown in Fig. \ref{fig_validation}.  Details 
           are discussed in the text.}
  \label{fig_variability}
\end{figure}

The variability of both models in 
Fig. \ref{fig_validation}, defined as the ratio 
between the standard deviation of all model 
observations from a 24-hour period and the 
24-hour average brightness from the same timespan, 
is shown in Fig. \ref{fig_variability}.  The variability 
steadily increases from about 5\% near full phase to 
about 30-40\% near crescent phase.  Brightness is more 
variable at crescent phases since the illuminated 
portion of the disk represents a relatively small 
fraction of the planet's surface area and, thus, is 
easily dominated by clouds that rotate into view or 
a cloud-free view of the glint spot.  Variability at 
gibbous phases following northern summer is slightly 
larger than variability at gibbous phases prior 
to northern summer (4\% versus 6\%, respectively), 
which is a seasonal effect.  The magnitude of the 
variability in our models agrees well with the 
simulations of \citet{oakley&cash09}.  However, 
these authors did not find a seasonal dependence 
in variability measurements.  Furthermore, their 
simulations show a sharp increase in variability 
as the planet moves into crescent phases 
(variability increases from 5-10\% to  
40\% over about 10$^{\circ}$ of orbital longitude) 
while our models show a gradual 
increase in variability into crescent phases.

\section{Discussion and Conclusions}

\subsection{Earth With and Without Glint}

Including phase-dependent reflection from oceans  
and clouds as well as Rayleigh scattering has allowed 
us to reproduce both the brightness and phase dependence 
of Earthshine observations. \citet{williams&gaidos08} 
explicitly ignored Rayleigh scattering and phase-dependent 
scattering from clouds and were unable to reproduce Earthshine 
observations, demonstrating the importance of including these 
effects in a realistic spectral 
Earth model.  A model that only includes 
phase-dependent Rayleigh scattering produces an  
increase in apparent albedo at crescent phases (due to 
weak forward and backward scattering lobes in 
the Rayleigh scattering phase function), but 
the upturn occurs only at extreme crescent phases (at 
orbital longitudes within 30$^{\circ}$ of new phase), which 
is not seen in Earthshine observations.  This argues 
that the lack of phase-dependent cloud scattering 
in the model presented in \citet{williams&gaidos08} was the 
primary reason why their model could not reproduce Earthshine 
observations.  Thus, predictions regarding the behavior of 
Earth's brightness at crescent phases are especially 
reliant on realistic cloud modeling.

Models that do not include the ``hiding" effect of 
ocean waves will over-estimate the brightness 
of water surfaces at glancing reflection angles.  For 
edge-on orbits (inclination, $i = 90^{\circ}$), 
this effect becomes especially important 
at orbital longitudes within about 30$^{\circ}$ of new 
phase, in agreement with  \citet{coxmunk54}.  The 
phase-dependent relative size of the glint spot, 
the tendency of water to be more reflective at glancing 
reflection angles, and the ``hiding" effect all combine 
to produce a maximum brightness excess for a realistic Earth 
over an Earth without glint near 30$^{\circ}$ from new phase 
(for an orbit viewed edge-on).  
Varying ocean wind speeds in our model 
show that the location of this peak is only weakly dependent 
on surface wind conditions.  Previous models used to 
investigate the detection of surface oceans 
\citep{williams&gaidos08,oakley&cash09} do not include the 
``hiding" effect and cannot make strong statements 
about glint detection at extreme crescent phases.

The season- and phase-dependent variability of Earth's 
brightness, shown in Fig.~\ref{fig_variability} and 
taken from the edge-on simulations shown in 
Fig,~\ref{fig_validation}, is due to contrast 
between highly reflective surfaces and surfaces with 
low reflectivity.  Following full phase, 
which corresponds to northern summer in our simulations, 
snow and sea ice in the northern polar region have 
been replaced by darker surfaces (\emph{e.g.,} grassland) 
which provide greater contrast to clouds as 
compared to the snow and ice present 
prior to full phase.  Thus, variability is larger following 
northern summer in our simulations.  At crescent 
phases, contrast is provided by bright, forward-scattering 
clouds, and/or the bright glint spot, against relatively 
dark, Lambertian-scattering surfaces.  The illuminated sliver of 
the planet at crescent phases represents a relatively 
small amount of surface area so that the illuminated 
disk at these phases can become dominated by large 
cloud features (or the glint spot), leading to large 
variability.  Near full phase, the illuminated disk 
represents a large amount of surface area, averaging 
over clouded and non-clouded scenes, leading to 
relatively low variability.

Our simulations without glint produce the 
same rise in variability into crescent phases as our 
simulations with glint, which indicates that 
variability at crescent phases is not completely driven 
by glint.  Thus, the trend of increasing 
variability into crescent phases 
is not a clear indicator of surface oceans, as was 
proposed by \citet{oakley&cash09}.  Any planet that 
can achieve sufficient contrast between bright and dark 
surfaces will produce a variability signal that 
increases into crescent phases, regardless of the presence 
of oceans.  


Our glinting model demonstrates a 
wavelength-dependent brightness excess over our 
non-glinting model, since some 
wavelengths are more sensitive to surface 
effects than others. The excess shrinks to less than 
10\% for the wavelength range 0.3-0.4 $\mu$m,  
where Rayleigh scattering obscures the surface.  
At wavelengths that correspond to 
relatively deep absorption bands, like the 
1.4 $\mu$m water band, the excess shrinks to 
nearly zero because observations are mostly  
insensitive to the surface.  

At crescent phases, pathlengths through the 
atmosphere are relatively large and optical 
depths to Rayleigh scattering can be larger than unity  
even at longer wavelengths.  This indicates 
that observations which aim to detect the brightness 
excess due to glint should be made at wavelengths 
in the near-infrared range.  Earth's brightness drops 
by over an order of magnitude between 1-2 $\mu$m, 
arguing that searches for glint 
should occur below 2 $\mu$m for higher signal-to-noise ratio 
(SNR) detections.  Since glint is a broad feature in 
wavelength space (it is the reflected solar spectrum, 
modulated by Rayleigh scattering, liquid water 
absorption at the surface, and atmospheric 
absorption), photometry can be used to detect glint 
provided that strong absorption features are 
avoided.

\subsection{Observing Requirements for Glint Detection}

The glint effect is strongest at NIR wavelengths, so 
that pairing JWST \citep{gardneretal06} with an external 
occulter \citep[e.g.,][]{soummeretal09,cash06} would 
present a near-term opportunity for the detection of 
oceans on extrasolar planets.  Here we discuss 
optimal filter selection and inner working angle 
(IWA) and SNR requirements for glint detection at 
wavelengths accessible to JWST.

The JWST Near Infrared Camera (NIRCam) 
\citep{horner&rieke04} offers several 
medium and wide band filters suitable for glint 
detection, with the F115W, F150W, and F162M filters 
(spanning 1.0-1.3 $\mu$m, 1.3-1.7 $\mu$m, and 
1.55-1.70 $\mu$m, respectively) being most ideal.  The 
F115W and F150W filters partially overlap water absorption 
features but experience a photon flux 3-4 times larger than 
the F162M filter, which does not span any strong water 
features and, thus, has an increased sensitivity to 
surface effects.

\begin{figure}[ht]
  \centering
   {
     \includegraphics[width=4in,angle=0,trim=0mm 22.5mm 0mm 22.5mm,clip]{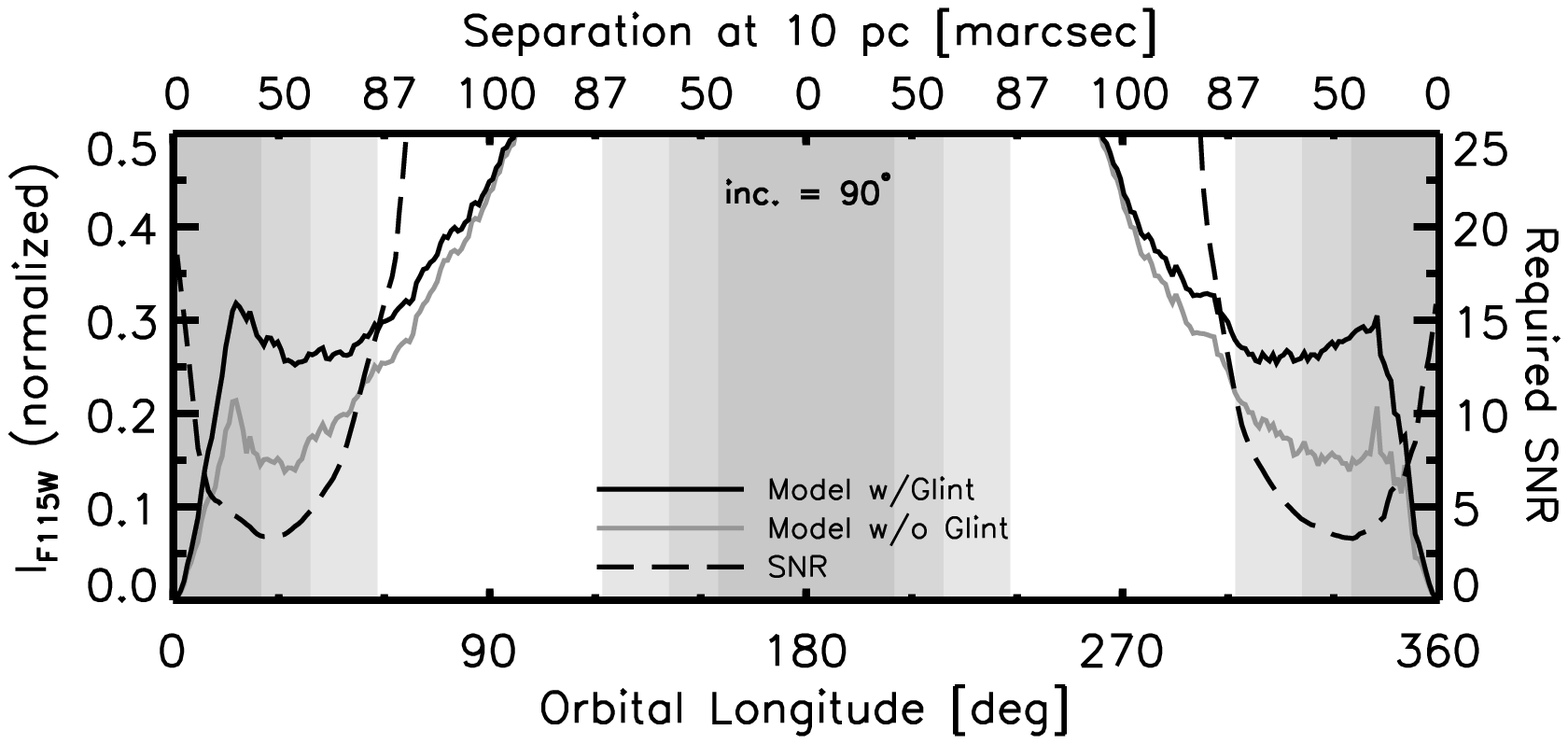}
   }
   { 
     \includegraphics[width=4in,angle=0,trim=0mm 22.5mm 0mm 22.5mm,clip]{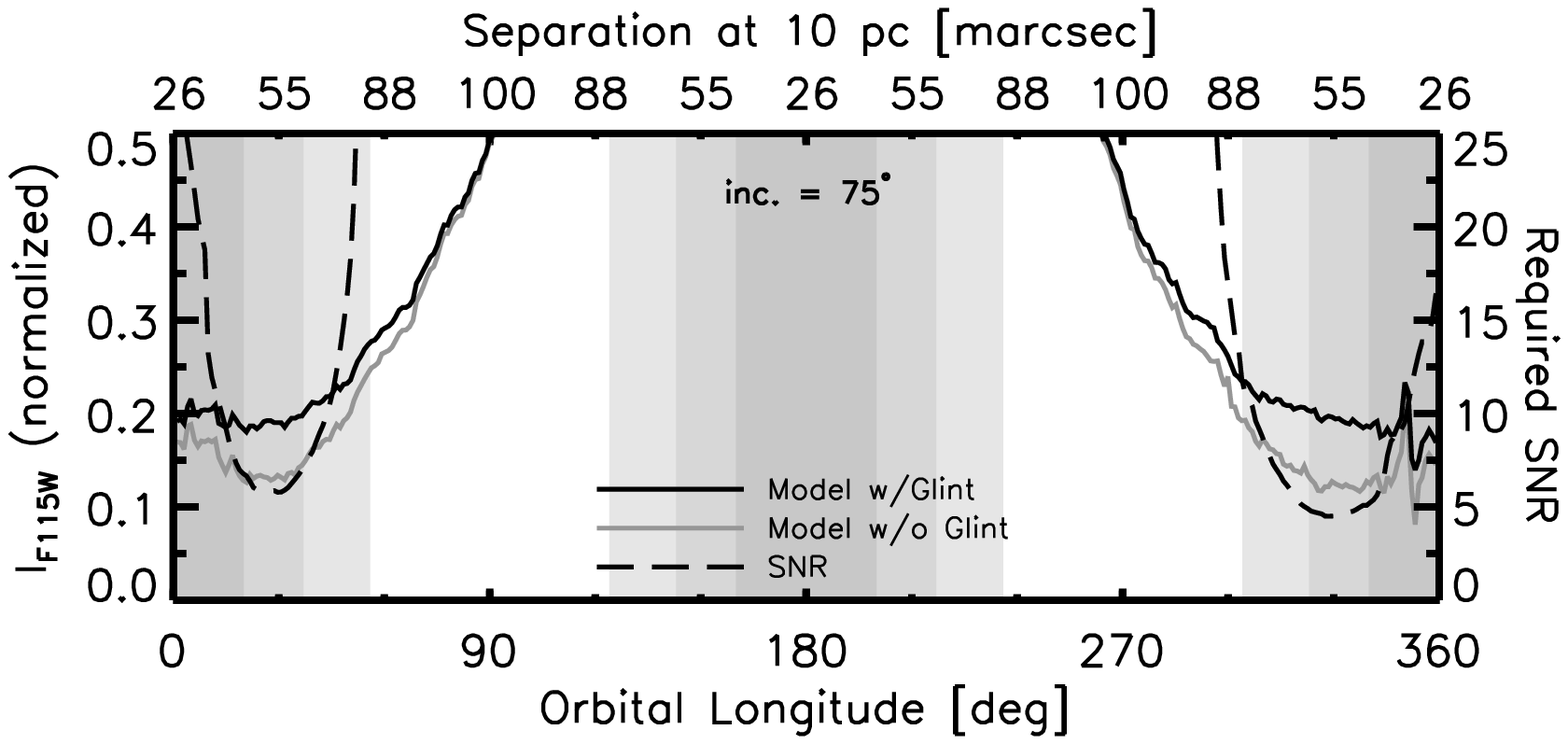}
   }
   { 
     \includegraphics[width=4in,angle=0,trim=0mm 22.5mm 0mm 22.5mm,clip]{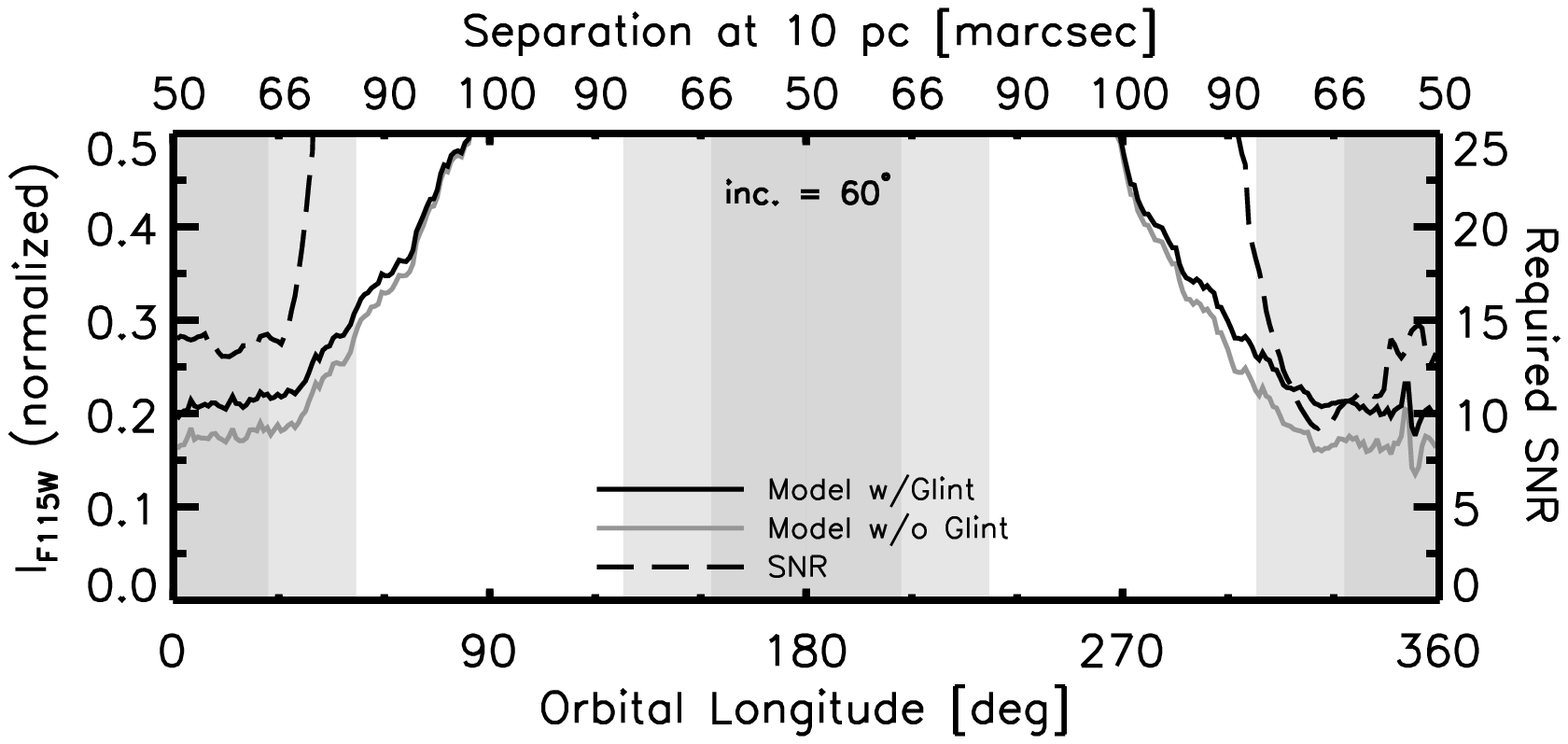}
   }

  \caption{\footnotesize
           Earth's brightness through the JWST/NIRCam F115W filter (spanning 
           1.0-1.3 $\mu$m) relative to its brightness at  
           gibbous phase (135$^{\circ}$ and 225$^{\circ}$ orbital 
           longitude) for orbital inclinations of 90$^{\circ}$ (top), 
           75$^{\circ}$ (middle), and 60$^{\circ}$ (bottom).  Glinting 
           model is in black and non-glinting model is in grey. Vertical 
           shaded regions indicate the portions of the orbit for which 
           a planet orbiting at 1 AU from its host star is within 
           85 milli-arcseconds, which is a standard IWA for an occulter 
           paired with JWST \citep{brown&soummer10}, for a system at a 
           distance of 5 parsecs (darkest), 7.5 parsecs (medium), and 
           10 parsecs (lightest).  Planet-star separation at a 
           distance of 10 parsecs is shown on the upper x-axis.  The 
           SNR required to distinguish the glinting model from the 
           non-glinting model at the 1-$\sigma$ level is shown along 
           the right y-axis and corresponds to the dashed line.
           Observations have been averaged over 24-hour periods.}
  \label{fig_jwst}
  \label{lastfig}
\end{figure}

Figure \ref{fig_jwst} demonstrates a strategy that 
could be used to observe the brightness excess 
from glint.  We show the 24-hour average brightness 
of Earth through the F115W NIRCam filter normalized to 
an observation at gibbous phase for several different 
orbital inclinations.  Also shown is the SNR required to 
distinguish our glinting model from our non-glinting 
model at the 1-$\sigma$ level assuming that high SNR 
observations (SNR=20) have been made at 
gibbous phase. For the inclination equals 90$^{\circ}$ and 
75$^{\circ}$ cases, the glinting model  
shows a distinct leveling-off of its normalized 
brightness at crescent phases.  Relative to its 
brightness at gibbous phase, the 
planet's brightness can remain roughly constant 
through certain phases due to the competing 
effects of the falling stellar illumination 
and the rising reflectivity 
due to glint.  The glint effect becomes difficult to 
detect for the case where inclination equals 60$^{\circ}$ 
because the planet never becomes 
a small enough crescent to produce a strong glint 
effect.  Minimum SNRs between 5-10 are required 
for glint detection, depending on viewing geometry and 
telescope IWA.  Note that roughly 25\% of all planets 
discovered will have orbital inclinations between 
75-105$^{\circ}$ and 50\% will have inclinations 
between 60-120$^{\circ}$.

Observations of Earth-like exoplanets at crescent 
phases can be difficult due to IWA constraints.  
Vertical shaded regions in Fig. \ref{fig_jwst} 
represent portions of the planet's 
orbit that cannot be observed for an Earth-twin at 
different distances assuming an IWA of 85 
milli-arcseconds \citep{brown&soummer10}, demonstrating 
that the glint effect could be detected for near edge-on 
orbits out to a distance of about 8 parsecs for this IWA.  
The measure of planet-star separation shown along the top 
of the sub-figures indicates that an IWA of about 50 
milli-arcseconds would allow for the easiest 
detection of the glint effect for planets 
within 10 parsecs.  Note that angular separation will 
scale inversely with distance to the system, and that 
the IWA constraints become significantly less strict 
for an Earth-like planet at the outer edge of the 
Habitable Zone.

\section{Conclusion}
Our model successfully reproduces Earth's phase-dependent 
brightness from Earthshine observations.  We have shown 
that glint increases Earth's brightness 
by as much as 100\% at crescent phases, and that this 
effect is strongest in wavelength regions that are 
unaffected by Rayleigh scattering and atmospheric 
absorption.  The glint effect may be detectable  
using JWST/NIRCam paired with an external occulter.  
Depending on viewing geometry, minimum SNRs 
of 5-10 are required for glint detection and 
an optimal IWA for detection is about 50 
milli-arcseconds for an Earth-twin at 10 parsecs.

\acknowledgments

This work was performed by the NASA Astrobiology Institute's 
Virtual Planetary Laboratory, supported under solicitation 
No.~NNH05ZDA001C.

\clearpage

\section{Tables and Figures}

\end{document}